\documentclass[conference]{IEEEtran}
\IEEEoverridecommandlockouts
\usepackage{cite}
\usepackage{amsmath,amssymb,amsfonts}
\usepackage{algorithmic}
\usepackage{graphicx}
\usepackage{textcomp}
\usepackage{xcolor}
\def\BibTeX{{\rm B\kern-.05em{\sc i\kern-.025em b}\kern-.08em
    T\kern-.1667em\lower.7ex\hbox{E}\kern-.125emX}}
\begin{document}

\title{Effect of Data Degradation on Motion Re-Identification
}

\author{

\IEEEauthorblockN{Vivek Nair}
\IEEEauthorblockA{\textit{Computer Science} \\
\textit{UC Berkeley}\\
Berkeley, USA \\
vcn@berkeley.edu}

\and
\IEEEauthorblockN{Mark Roman Miller}
\IEEEauthorblockA{\textit{Computer Science} \\
\textit{Illinois Institute of Technology}\\
Chicago, USA \\
mmiller30@iit.edu}

\and
\IEEEauthorblockN{Rui Wang}
\IEEEauthorblockA{
\textit{Carnegie Mellon University}\\
Pittsburgh, USA \\
ruiwang3@andrew.cmu.edu}

\and
\IEEEauthorblockN{Brandon Huang}
\IEEEauthorblockA{
\textit{UC Berkeley}\\
Berkeley, USA \\
zhaobin@berkeley.edu}

\and
\IEEEauthorblockN{Christian Rack}
\IEEEauthorblockA{\textit{Human-Computer Interaction} \\
\textit{University of Würzburg}\\
W{\"u}rzburg, Germany \\
christian.rack@uni-wuerzburg.de}

\and
\IEEEauthorblockN{Marc Erich Latoschik}
\IEEEauthorblockA{\textit{Human-Computer Interaction} \\
\textit{University of Würzburg}\\
W{\"u}rzburg, Germany \\
marc.latoschik@uni-wuerzburg.de}

\and
\IEEEauthorblockN{James F. O'Brien}
\IEEEauthorblockA{\textit{Computer Science} \\
\textit{UC Berkeley}\\
Berkeley, USA \\
job@berkeley.edu}

}

\maketitle

\begin{abstract}
The use of virtual and augmented reality devices is increasing, but these sensor-rich devices pose risks to privacy. The ability to track a user's motion and infer the identity or characteristics of the user poses a privacy risk that has received significant attention. Existing deep-network-based defenses against this risk, however, require significant amounts of training data and have not yet been shown to generalize beyond specific applications. In this work, we study the effect of signal degradation on identifiability, specifically through added noise, reduced framerate, reduced precision, and reduced dimensionality of the data. Our experiment shows that state-of-the-art identification attacks still achieve near-perfect accuracy for each of these degradations. This negative result demonstrates the difficulty of anonymizing this motion data and gives some justification to the existing data- and compute-intensive deep-network based methods.
\end{abstract}


\section{Introduction}

In recent years, both research, corporate, and consumer interest in virtual reality (VR) has been growing \cite{wang2023metaverse,falchuk_social_2018,di_pietro_metaverse_2021}. The growing adoption of VR devices raises the priority of understanding novel security and privacy threats possible with these devices' use. One class of threats is a privacy risk due to the collection of motion data, a kind of data that is fundamentally necessary for the operation of VR headsets. This data, usually consisting of the 3D position and orientation of the headset itself as well as the hand controllers, can be used to identify users \cite{Miller2020,nair_unique_2023,Miller2021,Miller2022}. On one hand, this behavioral biometric is a benefit to security, as device developers can verify the user of the device continuously and implicitly. On the other hand, our focus is the privacy risk of this motion data, either by malicious applications or by malicious other users in social VR.

Recent work has demonstrated inference \cite{miller_large-scale_2023, nair_inferring_2023} and identifiability \cite{Miller2020,Moore2021,schell_extensible_2023}, even in massive pools of people \cite{nair_unique_2023}. This has, in turn, motivated defenses \cite{Nair2022a,nair2023deep}. However, these defenses are either very data-intensive, requiring thousands of recordings from hundreds of users, or they do not protect against sample-level data attacks. A simpler approach based on data degradation would be more desirable than a compute- and data-intensive approach, if it were effective.

In this work, we explore the possibility of defending against this identification attack using various methods of reducing the quality of the data. Specifically, we consider:
\begin{itemize}
    \item \textit{sample-level noise,} the addition of Gaussian noise to each dimension of motion data
    \item \textit{reduced framerate,} the removal of intervening frames within recorded data by subsampling
    \item \textit{reduced precision,} rounding the data to the nearest subdivisions of a meter, and
    \item \textit{reduced dimensionality,} the reduction of the data available to the model to a single dimension.
\end{itemize}

However, we find identifying signals are robust to these types of degradation. This work demonstrates that the motion necessary to identify an individual is not limited to high-quality data or a controlled environment, but is robust in many situations. This robustness poses a greater challenge to privacy enhancing technologies.

\section{Related Work}

To set the context for the current work, we first review identification using motion. Then, we review the current space of defenses against identification.

\subsection{Identification using Motion}

Because the identification of a user by their motion is both the goal of trusted authentication systems and identification attacks, we interleave these two domains within this review. The distinction we draw in this space is whether the entity with the data (attacker or authenticator) has access to motion data \textit{only}, or whether they have access to some other aspect of the system. To give examples, these other aspects can be the intent of the mover, e.g., asking the user to perform a specific action; presence within the mover's world, e.g., social engineering attacks like waving and eliciting a wave back; or design of the virtual environment, e.g., eliciting certain actions when playing a virtual game.

Notably, while the focus of this work is on privacy, there is an alignment between the present work and the identification of cooperative users. Most of these identification methods are types of \textit{implicit authentication}, authentication based upon actions a user carries out for other reasons \cite{jakobsson2009implicit}, and \textit{continuous authentication}, in which users are authenticated several times throughout a session, addressing session hijacking \cite{Traor2012}. The same finding - that a particular set of actions and features can be identifying - is a risk to privacy and a benefit to security and usability on these fronts. 

In contrast to authentication, we are interested in adversarial identification, where a weaker adversary has access to only the motion data of the target. This kind of attack does not require the trust of the target, access to social interactions with the target \cite{Falk2021}, or access to the environment the target is in \cite{Nair2022}.

\subsection{Defenses}

In contrast to the attack space, there is less work about defense
mechanisms for this data. M. Miller and collaborators \cite{Miller2020}
reduce the training data streams from 18DOF (head and hands position
and rotation) to 3DOF (head rotation only) and reduce accuracy from
95\% to 20\% on a set of 511. Moore and collaborators \cite{Moore2021}
reduce accuracy from 89\% to 32\% on one set of data and 42\% to 13\%
on a second by switching from position-based to velocity-based feature
vectors. Nair, Gonzalo, and Song \cite{Nair2022a} use differential
privacy methods on the biometric features
they lay out in previous work \cite{Nair2022}. Differential privacy methods incorporate a type of noise to each data point within a dataset so that even when the entirety of the dataset is compromised save for one point, the relative likelihood between the data point being the true value and the data point being any other value is bounded above by the privacy parameter. For a formal mathematical definition, see \cite{dwork_algorithmic_2013}. However, because the protection is only applied to a handful of hand-selected features (height, arm span, etc), the protection leaves many kinds of identifiers unchanged (e.g., degree to which a user looks around a space). There are other types of privacy guarantees, such as $k$-anonymity and plausible deniability. While to our knowledge these approaches have not been taken on sample-level data, there has been work on protecting eye-tracking data \cite{david-john_privacy-preserving_2023}.

Leveraging more advanced techniques, Nair and collaborators also have proposed \textit{Deep Motion Masking} \cite{nair2023deep}, which breaks down motion using LSTMs into the variance due to the \textit{action} and variance due to the \textit{user} - in essence, subtracting out the idiosyncrasies of the individual before transmitting motion data. This reduces the identifiability of motions stream while maintaining plausibly-human behavior. However, this work requires significant data and compute power and may not extend to out-of-distribution actions. This motivates us to explore the potential effectiveness of simpler methods to de-identify data, specifically data obfuscation or degradation.

One work that is most similar to the approach here is work by Hanisch and collaborators \cite{hanisch_understanding_2023} who investigate the identifiability of gait subject to perturbations, coarsening, removal of data, and normalization. They ultimately conclude that gait anonymization is highly challenging, given their results that the anonymization techniques, for the most part, did not reduce accuracy. In contrast, we investigate Beat Saber, which has a significantly different macro-level structure (see the Data section for more information) and has only recently been discovered to be identifiable motion.

\section{Methods}

\subsection{Threat model}

It is important to establish the kind of threat under study. In previous
work on VR and identifiability, there are two dimensions upon which
researchers have categorized threats. First, there is the question
of what data is available to the attacker. Nair and collaborators
\cite{Nair2022} delineate between hardware-level attackers that have
access to firmware, client-level attackers that have access to the
headset APIs, server-level attackers that have access to the telemetry
data sent to the servers and 'unprivileged user' attacker which is
another VR system partaking in the same social virtual world. Along
this dimension, we focus on the unprivileged user.

The second aspect of space of threat models is the capability of the
attacker to influence the behavior of the participant, and the extent
to which this can be done. For example, is the attacker designing
a virtual world \cite{Nair2022}, are they another user that is interacting
with the target \cite{Falk2021}, or do they wish not to interact with
the target entirely? In our work, we focus on no interaction at all.
This may occur because the attacker is working with previously-collected
data, does not want to be vulnerable in the virtual world, or has data
collected at scale and cannot interact with each target.

Per the framework of Garrido et al. \cite{garrido2023sok}, the adversary of interest to us is the ``user adversary.''
This threat actor is selected because it is the least privileged attacker.
Therefore, findings based on this work are likely to be applicable
to all attacks leveraging VR pose tracking data. It also sets a baseline on
threat for all these other conditions. Finally, there are some cases
in which this may be the mode of an attacker, e.g., large-scale surveillance
where individuals are not queried directly, re-identification attacks
where actions are stored for a period of time before being queried,
or any other situations in which the attacker does not which to have
any direct interaction with the target. Note that this threat model
is quite different from the traditional authentication threat model
in which a user attempts to gain unauthorized access by posing as
another user.

\subsection{Data}

The data used in this work comes from the Berkeley Open Extended Reality Recordings 2023 (BOXRR-23) dataset \cite{nair2023berkeley}.
BOXRR-23 consists of 4.7 million motion capture recordings from 105,852 users, derived from ``Beat Saber,'' a popular virtual reality rhythm game, and ``Tilt Brush,'' a virtual reality drawing application. In this paper, we only use the 500 users with the most recordings from the BOXRR-23 dataset. For each of these users, at least 500 separate recordings are present, with sessions varying in length. After sorting the recordings chronologically, the first 400 recordings per user are used for training, the next 50 are used for validation, and the final 50 are used for testing.

\subsection{Feature Engineering}


The registration of a coordinate system is often not amenable to moving,
flexible, and diverse human bodies. Over time, different coordinate
systems have been developed for specific purposes, such as the anatomical
planes (coronal, sagittal, transverse) for medical terminology. For
the purposes of our work specifically and of VR more generally, we
use a coordinate system that synthesizes the global vertical axis
with horizontal axes relative to the headset's forward direction, known as \textit{body-relative coordinates} \cite{schell_extensible_2023, 10024474}.

To perform this normalization, the forward direction of the head (headset) is projected onto the horizontal plane. The transformation applied to all tracked objects (left hand controller, right hand controller) is the inverse rotation about the vertical axis so that the projected forward direction of the head aligns with the forward direction of the coordinate system.
In regards to the question at hand,
this would mean the body-space coordinate system is likely to be more
effective at separating one's pose from another's than the global
coordinate system would be. 
The use of body-relative coordinates for VR identification models is equivalent to that proposed by Rack et al. \cite{schell_extensible_2023} and is enabled by the Motion Learning Toolkit \cite{2023_MLT}.


For the features, at each frame processed by the VR device, the position and orientation of the user's left hand, right hand, and head are captured. Three positional coordinates and four orientation coordinates (in quaternion format) are captured for each of the three tracked objects, totaling 21 dimensions captured per frame. After the body-relative transformation is applied, 18 dimensions remain, as the three positional coordinates of the head are eliminated by this transformation. The vertical rotation of the head is also eliminated, but the quaternion representation of the rotation retains use of all four dimensions. The values of interest to us are the first and second derivatives of these 18 values; the result is 36 values per frame describing \textit{body-relative velocity} and \textit{body-relative acceleration}.

Each user's VR device may render frames at a slightly different frequency due to a variety of external factors. To eliminate frame rate as a potential confounding factor, we first resample all motion capture streams to a constant 30 frames per second by using a linear interpolation for positional coordinates and a spherical linear interpolation for orientation quaternions. Each session of a user was then split into 30-second sequences. In summary, an individual sequence has 30 seconds, 30 frames a second, and 36 values per frame; thus, our model has an input shape of $(900 \times 36)$. 

\subsection{Model}
\label{sec:architecture}

The model's task is to identify a user based upon their motion. More formally, the model is given a $(900 \times 36)$ sequence as described above. With that sequence, the model attempts to predict the participant who generated that motion, represented as a value of a categorical variable encoded with a one-hot encoding.

The model we have selected is a Long Short-Term Memory (LSTM) model \cite{hochreiter1997long}, implemented in Python version 3.10.2 using Keras version 2.10.1.
The choice of LSTM was to take advantage of the sequential nature of the data. Most hyperparameters for the model were left to the defaults; in particular, the Adam optimizer \cite{kingma2014adam} was used with a learning rate of $0.001$. Specifically, we utilize the ``LSTM Funnel'' architecture described by Nair et al. \cite{nair2023deep}.

The predictions were made per session by taking the entire session
of pose tracking data, computing 30-second sequences as described above, and then summing the logarithmic probability of each user reported by the model across all samples. We interpreted this distribution as a probability estimation for the classification of the session
as a whole, in line with previous work \cite{Miller2020}.

The problem type is classification rather than a ranking problem, along the lines of similar work \cite{Miller2020,Moore2021,Olade2020}. This method can be contrasted with multiclass AUC, which increases with any improvement in identifiability, not just when a sample is correctly classified.

\section{Results}
We study the identifiability of motion data alone through four kinds of methods. As described below, we degraded the quality of the motion data signal in a variety of ways to evaluate what effect, if any, this had on the identification accuracy.

\subsection{Added Noise}

\begin{figure}[h]
    \centering
    \includegraphics[width=\linewidth]{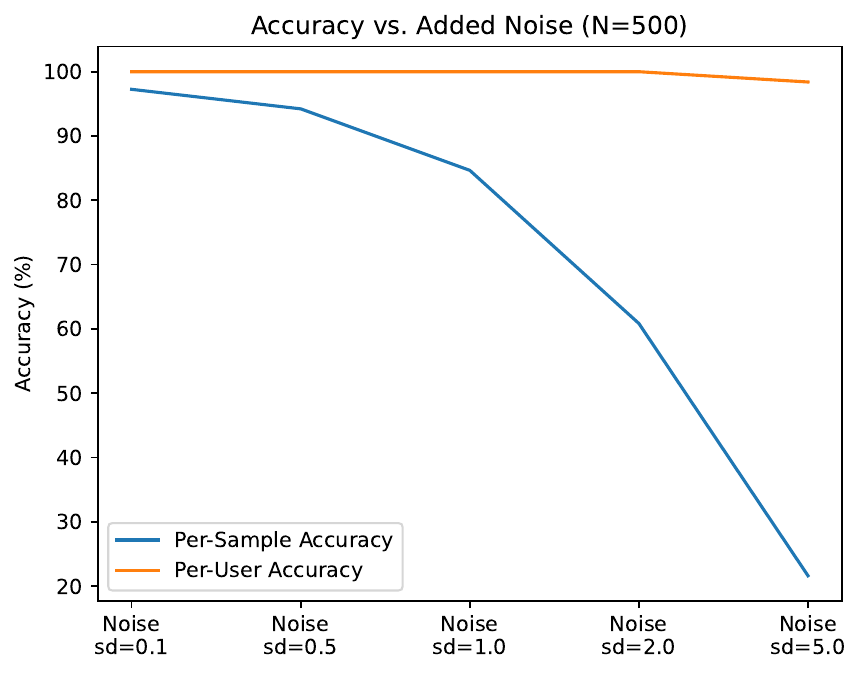}
    \caption{Cross-session identification accuracy with increasing Gaussian noise added to the telemetry signal}
    \label{fig:added-noise}
\end{figure}

First, we attempted to thwart the identification models by introducing random noise to each dimension of the motion telemetry stream. Specifically, we added zero-centered Gaussian noise with increasingly large standard deviations ($\sigma$) ranging from 0.1 to 5.0. This range represents a spread of values that interpolate between small, perceptible changes (10cm) to unrealistically large (5m). The results of this process are shown in Figure \ref{fig:added-noise}. We found that a per-user identification accuracy of 100\% can be achieved with noise as high as $\sigma=2.0$, and with over 90\% accuracy even when $\sigma=5.0$.

\subsection{Reduced FPS}

\begin{figure}[h]
    \centering
    \includegraphics[width=\linewidth]{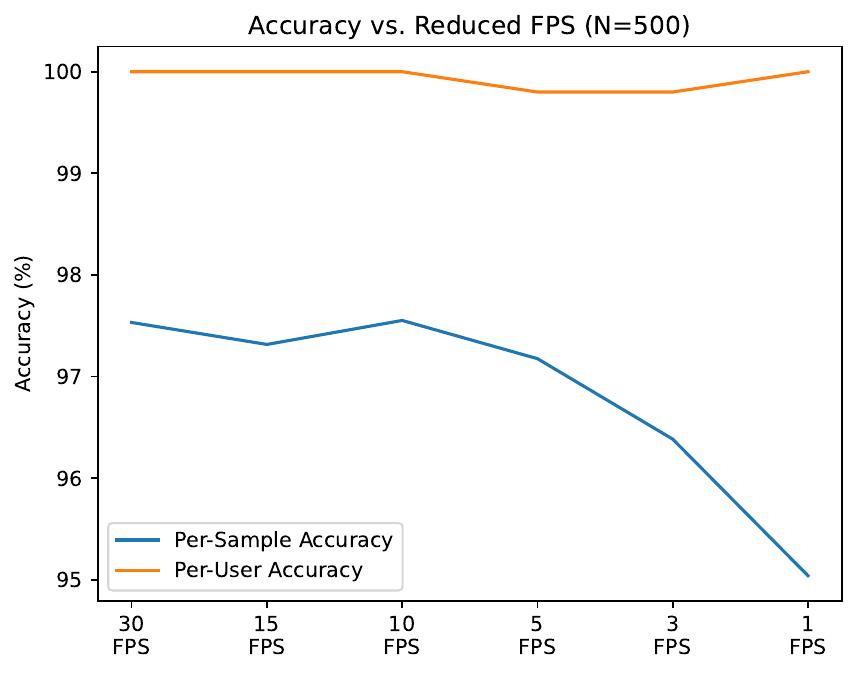}
    \caption{VR identification accuracy with reduced FPS}
    \label{fig:reduced-fps}
\end{figure}

Next, we attempted to reduce the frame rate from the baseline of 30 FPS to as low as 1 FPS. The method of reduction was direct subsampling (rather than interpolation). Interestingly, downsampling motion data to 15 FPS and 10 FPS resulted in almost no reduction in per-sample or per-user accuracy. Further reductions to 5 FPS, 3 FPS, and 1 FPS resulted in degraded per-sample accuracy, but 100\% per-user accuracy was still achieved at just 1 FPS, as shown in Figure \ref{fig:reduced-fps}.

\subsection{Reduced Precision}

\begin{figure}[h]
    \centering
    \includegraphics[width=\linewidth]{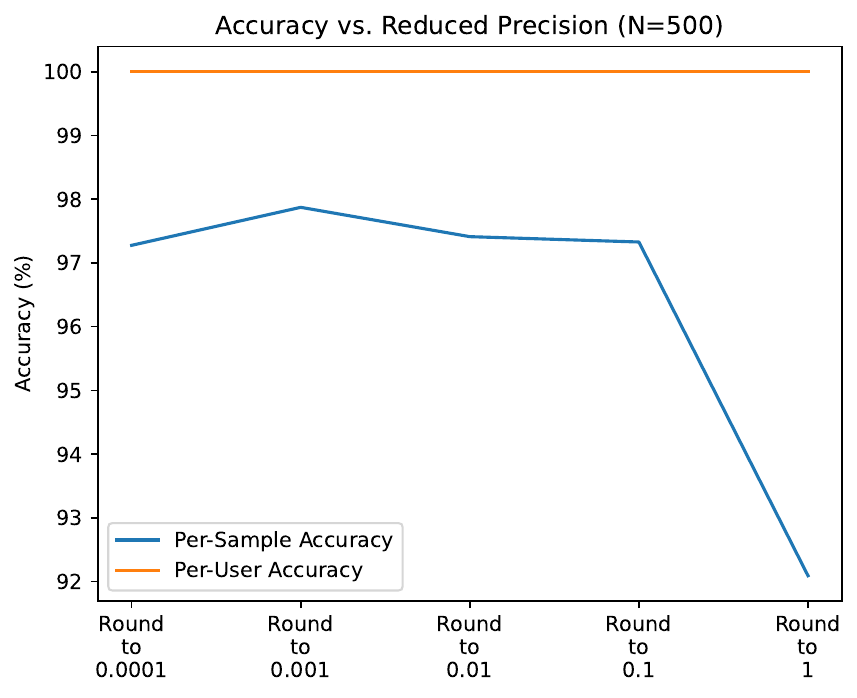}
    \caption{VR identification accuracy with rounded signal}
    \label{fig:reduced-precision}
\end{figure}

We also attempted to reduce the precision of VR motion data in the spatial (rather than temporal) dimensions. We did so by rounding all values in the VR motion data to the nearest 0.0001, 0.001, 0.01, 0.1, and 1 meters. The results, shown in Figure \ref{fig:reduced-precision}, show minimal impact on accuracy for all degrees of rounding between 0.0001 and 0.1. Rounding all dimensions to the nearest full meter did have a significant impact on per-sample accuracy, but still allowed 100\% per-user accuracy to be achieved.

\subsection{Reduced Dimensions}

\begin{figure}[h]
    \centering
    \includegraphics[width=\linewidth]{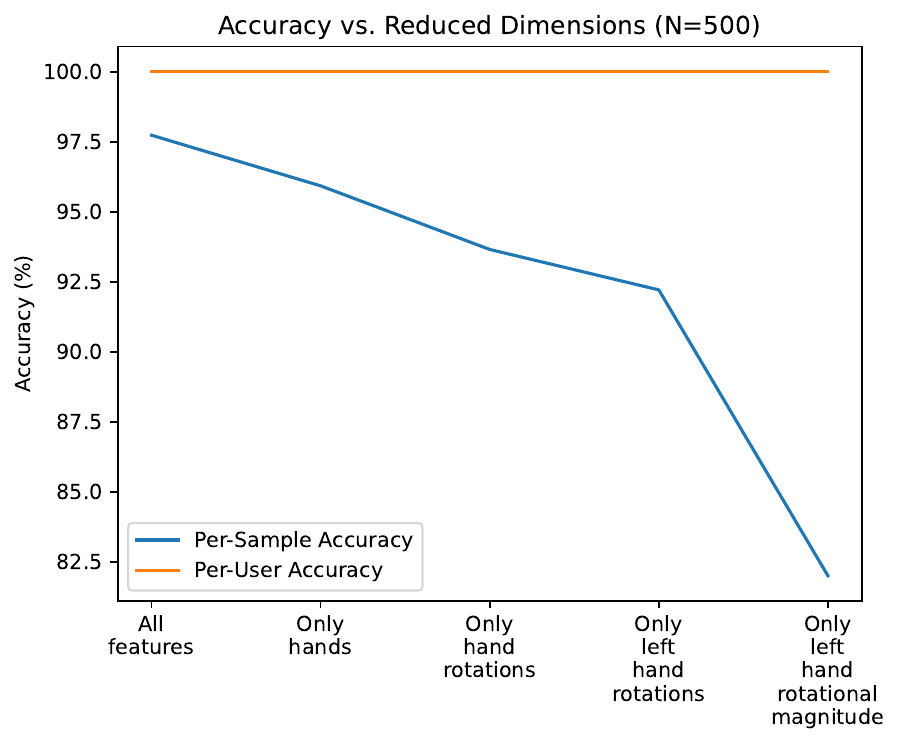}
    \caption{VR identification accuracy with reduced dimensions}
    \label{fig:reduced-dimensions}
\end{figure}

Finally, we attempted to reduce the dimensionality of the motion data. We began by eliminating the dimensions associated with the users' heads, leaving only their hands. Next, we eliminated all positional dimensions, leaving only the rotations of the users' hands. Further, we eliminated the users' right hands, leaving only their left hand rotations. Finally, we eliminated the i, j, and k quaternion elements, leaving only the w coordinate of the quaternion, which corresponds to rotational magnitude. The results of these reductions are shown in Figure \ref{fig:reduced-dimensions}; each dimensionality reduction accompanied a corresponding drop in per-sample accuracy, with 100\% per-user accuracy still being observed.

\section{Discussion}

\subsection{Summary of Results}

We performed several methods of signal degradation on VR motion data. In each case, whether Gaussian noise is added to the data, the framerate is reduced, the values are rounded, or only one dimension is tracked, identification is still very high, leading to an accuracy of 100\% at the session level. This illustrates that given enough data, it does not take ideal conditions for motion data to be identifying. Rather, based upon these results, it would seem a user is more likely to run into user experience issues through the obfuscation than to successfully anonymize the data through signal degradation. For example, negotiating personal space in a social VR setting would be very difficult to do if other avatar positions are only updated once per second.

\subsection{Implications for Privacy}

First, the degraded signal results further dispel the myth that motion-based identification in VR requires high-quality data or a controlled laboratory environment. In fact, we show that a low-quality device or network will still be good enough to identify VR because the signal - whatever is identifying within the motion - is very robust. In short, you don't need many conditions for identification to be feasible, just a lot of data.

Considering virtual reality motion identification as a whole, there are several steps to take.
First, developers should protect this data with standard practices for personally identifying
data \cite{wang2023metaverse}. When this data needs to be shared with others, it can be helpful
to reduce the time span available, minimize variation in activities,
or modify data to produce security guarantees like $k$-anonymity, plausible deniability, or differential privacy \cite{Nair2022a,david-john_privacy-preserving_2023}.
Furthermore, there are developments in law that need to be made to
clarify the legal status of this data based on its risks to privacy
\cite{heller_watching_nodate}.

\subsection{Limitations and Future Work}

Some limitations of this work are that the manipulations are not applied together, e.g., there is no combination of added noise and reduced FPS. Additionally, this model is only trained on Beat Saber data, and while this application was not specially selected for its identifiability, it remains to be seen how this identifiability and potential defense extends beyond a single application type.

While some work \cite{Nair2022a} weakens the relationship between real-world
biometrics like height and arm length from a user's virtual avatar,
it may be plausible, given a user's preferences, to disconnect those two 
entirely. Care must be taken in this approach, though, as normalization may make other idiosyncrasies more prominent \cite{Liebers2021}. Another approach to avoid this tradeoff is to use transformed social interaction \cite{Bailenson2004}
so that gestures that might otherwise be identifiable can come from
another recorded value but still be communicative.

\section{Conclusion}

In this work, we test the effectiveness of signal degradation against state-of-the-art re-identification methods. Despite employing various degradation methods such as adding noise, reducing framerate, precision, or dimensionality, we found that identification accuracy remained remarkably high, almost always achieving 100\% at the session level. This result underscores the robustness of re-identification attacks based upon motion data. If simpler privacy protection methods are effective, they need to extend beyond these kinds of signal degradation; if simpler methods are not available, then the evidence can show the necessity of more complex machine learning to protect privacy. We hope this negative result can further define the boundaries between and aid future research in protecting privacy in VR.


\bibliographystyle{IEEEtran}
\bibliography{biblio}

\end{document}